\documentstyle[aps,preprint]{revtex}
\begin{document}
\title{Cross-over between first-order and critical wetting
at the liquid-vapour interface of n-alkane/methanol mixtures:\\
tricritical wetting and critical prewetting}
\author{A.I. Posazhennikova$^1$, J.O. Indekeu$^1$,
D. Ross$^2$, D. Bonn$^3$, and J. Meunier$^3$
\\
$^{1}$Laboratorium voor Vaste-Stoffysica en Magnetisme\\ Katholieke
Universiteit Leuven, B-3001 Leuven, Belgium\\ $^{2}$Process
Measurements, National Institute of Standards and Technology,
Gaithersburg, Maryland 20899, USA
\\ $^{3}$Laboratoire de Physique Statistique, Ecole Normale Sup\'erieure,
F-75231 Paris cedex 05, France }
\maketitle
\begin{abstract}
A simple mean-field theory is presented which describes the basic
observations of recent experiments revealing rich wetting behaviour
of n-alkane/methanol mixtures at the liquid-vapour interface. The
theory, qualitative and in part heuristic, is based on a microscopic
lattice-gas model from which a Cahn-Landau approach is distilled.
Besides the physics associated with the short-range components of
the intermolecular interactions, effects of the long-range tails of
the net van der Waals forces between interfaces are also taken into
account. Further, gravitational thinning of the wetting phase is
incorporated. The calculation of the spreading coefficient $S$ is
extended to the experimentally relevant situation in which the bulk
adsorbate is slightly away from two-phase coexistence due to
gravity. Analysis of this novel approximation to $S$ for systems
with short-range forces leads to the conclusion that the surface
specific heat exponents $\alpha_s = 1, 1/2, $ and 0, for first-order
wetting, tricritical wetting and critical wetting, respectively, are
robust with respect to (weak) gravitational thinning, consistently
with experiment. For three different systems the adsorption is
calculated as a function of temperature and compared with the
experimentally measured ellipticity. Including weak long-range
forces which favour wetting in the theory does not visibly alter the
critical wetting transition for the nonane/methanol mixture, in
contrast with the generic expectation of first-order wetting for
such systems, but in good agreement with experiment. For
decane/methanol weak long-range forces bring the transition very
close to the prewetting critical point, leading to an adsorption
behaviour closely reminiscent of short-range tricritical wetting,
observed experimentally for alkane chain length between 9.6 and 10.
Finally, for undecane/methanol the transition is clearly of first
order. First-order wetting is also seen in the experiment.
\end{abstract}
\newpage

\setcounter{equation}{0}
\renewcommand{\theequation}{\thesection.\arabic{equation}}
\section{Introduction and perspective}
Recent experiments have shown that a binary liquid mixture of linear
or ``normal" alkane and methanol in equilibrium with their common
vapour displays a first-order wetting transition if the wetting
temperature $T_w$ is well below the consolute-point temperature
$T_c$ and a ``short-range critical wetting" transition if $T_w$ is
very close to $T_c$~\cite{Ross1,Ross2}. The latter is the case for
n-nonane and methanol. The ``substrate" in these wetting experiments
is the saturated vapour phase. A recent review covers various
examples of experimentally observed first-order or continuous
wetting transitions in liquid mixtures~\cite{Bonnrev}.

The observation of short-range critical wetting in adsorbed binary
liquid mixtures is surprising, to say the least. Indeed, due to the
presence of van der Waals forces, which induce an algebraically
decaying long-ranged surface-interface interaction favouring
wetting, a first-order wetting transition should be
expected~\cite{deG1,NSS,P,NI,KM,ES}. Alternatively, if the van der
Waals forces oppose wetting, no wetting transition should occur,
unless the leading van der Waals interaction amplitude changes sign
at some temperature, resulting in ``long-range critical
wetting"~\cite{KM,DS,ESS,ES,GD}, observed experimentally in pentane
on water~\cite{RMBIB}. However, within the experimentally accessible
range of small film thicknesses (up to $100 \AA$) the van der Waals
forces can be neglected compared to exponentially decaying
mean-field (MF) and fluctuation-induced (FI)
interactions~\cite{Ross1,Ross2}. The range of the MF and FI
interactions is determined by the bulk correlation length, which can
become as large as the wavelength of visible light very close to the
upper consolute point, at $T_{c}$, where the two fluid phases become
identical.

The experiments, and in particular the measured value of the surface
specific-heat exponent, agree with the predictions of the mean-field
theory, and disagree strongly with renormalization-group predictions
based on a capillary-wave model~\cite{BHL}$-$~\cite{B}. Furthermore,
Monte Carlo simulations of short-range critical wetting~\cite{BLK}
also disagree with the RG predictions but agree with the
experiments. This behaviour can be understood in the light of more
sophisticated interface models~\cite{BP,PB,PS} which indicate that
the critical region around the transition, in which deviations from
MF behaviour show up for the surface specific heat exponent, is too
small to be relevant for either experiment or simulation. In other
words, mean-field theory should be an excellent description of
simulations and experiments of critical wetting with short-ranged
forces. The experimental observations of mean-field like behaviour
are thus entirely in keeping with the latest RG work, since in
particular the mass density difference between the wetting phase and
the surrounding bulk phase stop the parallel correlation length from
getting big enough to see fluctuation effects. We shall return in
detail to the relevance of this mass density difference in what
follows.

Our aim in this paper is to give a theoretical description of the
cross-over between the regimes of first-order wetting and
``short-range critical wetting" in this system, {\em which for
purely short-range forces would occur via a tricritical wetting
point}~\cite{NF}. In view of the observed consistency between
experiment and MF theory, and the predictions for the width of the
critical region discussed above, we adopt the point of view of the
classical theory of Cahn-Landau type. We include the van der Waals
forces as a weak perturbation and also take into account the
gravitational thinning of the wetting layer due to the difference in
the mass densities of the liquid phases. We neglect thermal
fluctuation effects, but incorporate the influence of the vicinity
of bulk criticality at the MF level. That is, the divergence of the
bulk correlation length is included in the theory, but with the MF
value for the critical exponent. In this way, the interplay of
wetting and critical adsorption is allowed for.

This paper is organized as follows. In section II we present the
microscopic lattice-gas model for the alkane/methanol/vapour system
and derive the Ising model coupling constants and fields from the
intermolecular and chemical potentials. In section III we extract
the continuum Cahn-Landau theory from the Ising model near the bulk
critical point. In section IV we study wetting layer thicknesses,
adsorptions, surface free energies, spreading coefficients and
critical exponents for the regimes of first-order, tricritical and
critical wetting within the model featuring only short range forces.
The cross-over from first-order to critical wetting including
long-range forces in the theory is investigated in Section V. There
we calculate the system parameters corresponding to nonane/methanol,
decane/methanol and undecane/methanol and derive adsorptions and
spreading coefficients assuming the long-range forces to be a weak
perturbation. We compare our results with the experiments and
investigate whether we can interpret them in terms of short-range
tricritical wetting or, alternatively, in terms of prewetting
criticality induced by long-range forces. In Section VI we present
our conclusions.

\setcounter{equation}{0}
\renewcommand{\theequation}{\thesection.\arabic{equation}}
\section{Microscopic lattice-gas model}
In this Section we adopt the philosophy of lattice-gas modeling
which is often applied to binary alloys etc., as exemplified in the
lectures of Yeomans~\cite{Y}. We consider a nearest-neighbour spin-1
Ising model on a 3-dimensional simple cubic (SC) or face-centered
cubic (FCC) lattice. The spin variable takes the value +1 (methanol
molecule), $-$1 (alkane molecule) or 0 (vacancy). The methanol-rich
phase sits at the bottom of the recipient, the alkane-rich phase is
in the middle, and the vapour is on top.

Given that our entire approach is based at mean-field level it would
be appropriate to start from a density-functional theory of a binary
mixture with long-ranged fluid-fluid forces, as can be derived
starting from this spin-1 model following the works of Dietrich and
Latz~\cite{DL} and Getta and Dietrich~\cite{GD}. However, the
application of this theory to non-spherical molecules such as
alkanes, and to polar molecules such as methanol, would only be a
first approximation, so that the difficult calculations inherent in
this approach would still not be sufficiently reliable. We therefore
feel the necessity to propose a much simpler strategy to get a
handle on the cross-over from first-order to continuous wetting in
the presence of long-range forces.

We introduce pair interaction energies $\epsilon_{MM}$ for
methanol-methanol, $\epsilon_{AA}$ for alkane-alkane, and
$\epsilon_{AM}$ for alkane-methanol pairs at nearest-neighbour
distance. The AA energy corresponds in a first approximation to the
Lennard-Jones potential well depth for the nonpolar alkane molecules
and the MM energy can be given either by the Lennard-Jones or, more
appropriately, by the Stockmayer potential well depth for the polar
methanol molecules~\cite{RS}. Furthermore, we introduce the chemical
potentials $\mu_M$ and $\mu_A$ for methanol and alkane particles,
respectively. We make the rough approximation that a {\em single
lattice constant} $\sigma$ suffices for the model, while in reality
the distances of closest approach, reflected e.g. by Lennard-Jones
diameters $\sigma_{MM}$ and $\sigma_{AA}$ can differ for methanol
and alkane molecules. Further, we ignore the chain conformation of
the n-alkanes and treat them effectively as {\em spheres}.

The spin-1 Ising model Hamiltonian reads
\begin{equation}
{\cal H}({s}) = -J \sum_{<ij>} s_i s_j - H \sum_{i} s_i - \Delta
\sum _{<ij>} (s_i s_j^2 + s_j s_i^2) - E \sum_{<ij>} s_i^2s_j^2 - M
\sum_{i} s_i^2,
\end{equation}
where the square brackets $<ij>$ indicate that the sums are over
nearest neighbours on the lattice. The Ising couplings can easily be
determined from the pair energies and chemical potentials,
considering the following pairs: vacancy-vacancy, methanol-vacancy,
alkane-vacancy, methanol-methanol, alkane-alkane and
alkane-methanol. This leads to the relations
\begin{eqnarray}
J & = & (\epsilon_{MM} + \epsilon_{AA} - 2 \epsilon_{AM})/4
\\
H & = & (\mu_M-\mu_A)/2 \\ \Delta & = & (\epsilon_{MM} - \epsilon
_{AA})/4 \\ E & = & (\epsilon_{MM} + \epsilon_{AA} + 2
\epsilon_{AM})/4 \\ M & = & (\mu_A + \mu_M)/2
\end{eqnarray}

Since the vapour phase is dilute and the liquid phases are dense, we
make the very reasonable approximation that the liquid is free of
vacancies and that the liquid-vapour interface is sharp. This allows
us to map the spin-1 model onto a spin-1/2 model {\em with a free
surface}. The vapour phase is thus replaced by an inert spectator
phase and vacancies no longer play a role. Therefore, inside the
adsorbate we have $s_i^2$ = 1 everywhere, and our model reduces to
one with spin 1/2 and bulk Hamiltonian
\begin{equation}
{\cal H}_{bulk} = -J \sum_{<ij>} s_i s_j - H \sum _i s_i - \Delta
\sum _{<ij>} (s_i + s_j) + \mbox{constant}
\end{equation}
For a lattice with coordination number $q$ ($q$ = 6 for SC and 12
for FCC) we can rewrite this and drop the irrelevant constant, so
that
\begin{equation}
{\cal H}_{bulk} =  -J \sum_{<ij>} s_i s_j - H_{bulk} \sum _i s_i
\end{equation}
where the bulk field is given by
\begin{equation}
H_{bulk} = H + q \Delta
\end{equation}
Bulk two-phase coexistence between the alkane-rich and methanol-rich
phases is reached for $H_{bulk}=0$ in this model.

We now proceed to derive the surface contribution to the
Hamiltonian, ${\cal H}_{surf}$. The surface layer of spins is
different from layers in the bulk in that there are missing bonds.
For the SC lattice there is 1 nearest neighbour missing per surface
site and for FCC there are 4 missing bonds, assuming a (100) surface
for simplicity. For a (111) surface there would be 3 missing bonds.
This leads to the result, for the surface layer,
\begin{equation}
{\cal H}_{surf} = -J \sum_{<ij>} s_is_j - H_{surf} \sum _i s_i
\end{equation}
where the surface field $H_{surf}$ takes the form
\begin{equation}
H_{surf} = H_{bulk}- m \Delta
\end{equation}
with $m=1$ for the SC lattice, and $m=4$ for the FCC lattice. At, or
very close to, bulk coexistence $H_{bulk} \approx 0$ so that the
surface field is governed by $ \Delta$. In particular, the surface
preferentially adsorbs that species for which the pair binding
energy is smallest (in absolute value), in order to minimize the
energy increase due to broken bonds.

We can estimate the leading surface field contribution $\Delta$ as
follows. For n-alkanes we inspect the liquid-gas critical
temperatures $T_c^{LG}$ and adopt the simple rule $\epsilon_{AA} =
0.75 k_BT_c^{LG}$ for obtaining the Lennard-Jones potential well
depth~\cite{RS}. This leads to the following estimates, from
n-pentane ($C_5H_{12}$) to n-undecane ($C_{11}H_{24}$):
$\epsilon_{AA}/k_B$= 353K (pentane), 380K (hexane), 405K (heptane),
427K (octane), 446K (nonane), 464K (decane), 480K (undecane). For
the polar molecule methanol, either we can use the Lennard-Jones
parameter derived using the same rule, which leads to
$\epsilon_{MM}/k_B$ = 385 K, or we can employ the Stockmayer
potential parameter $\epsilon^{S}_{MM}/k_B $= 417 K~\cite{RS}, which
is more suitable for polar molecules. Using (II.4) we see that {\em
the surface field changes sign} between hexane and heptane, when the
Lennard-Jones parameter is used for methanol. On the other hand, the
sign reversal of $\Delta$ occurs {\em between heptane and octane}
when the Stockmayer parameter is adopted. Experimentally, for short
chain length of alkane the alkane-rich phase is preferentially
adsorbed at the vapour phase, while for long chain length the
methanol-rich phase is preferred~\cite{Ross1,Ross2}. This is in
agreement with the microscopic model, which predicts $\Delta > 0$
for short alkanes. Experimentally, the reversal of preferential
adsorption takes place (approximately) for octane, which agrees well
with the theoretical prediction based on the  Stockmayer potential
parameter for methanol, which we will therefore use from now on.

\setcounter{equation}{0}
\renewcommand{\theequation}{\thesection.\arabic{equation}}
\section{Continuum Cahn-Landau theory}
In this section we apply the mean-field approximation to the Ising
Hamiltonian derived in the previous section, and subsequently make
the continuum approximation to obtain the Cahn-Landau theory. We
follow closely the derivation of Maritan, Langie and
Indekeu~\cite{MLI}, valid for slowly varying concentration profiles
and for temperatures close to the consolute point of the
alkane/methanol mixture. Additionally, we take into account that the
mixture is at, or close to, two-phase coexistence in  bulk.

Let $N$ be the number of cells in the lattice-gas representation of
our system and $F$ the free energy. We have $N = V/\sigma^3$, where
$V$ is the volume and $\sigma$ is the representative molecular
diameter, which serves as the lattice constant. The temperature at
the critical point of the binary liquid mixture, or {\em upper
consolute point}, is denoted by $T_c$. We consider the reduced free
energy density $f = F/(Nk_BT_c)$ as a function of the order
parameter $\phi$. In our system $\phi$ is the concentration
difference of methanol and alkane in the mixture, $x_M-x_A$, minus
the critical concentration difference, $x_{M,c}-x_{A,c}$. We have,
from $F=U-TS$,
\begin{equation}
f(\phi) = u(\phi) - T \sigma (\phi)/T_c
\end{equation}
with energy density
\begin{equation}
u(\phi) = - \phi^2/2 -h \phi + \mbox{constant}
\end{equation}
where $h$ stands for the reduced bulk field, $h=H_{bulk}/k_BT_c$.
The entropy density~\cite{MLI}, to 4th order in $\phi$, is given by
\begin{equation}
\sigma (\phi) = \ln 2 -\phi^2/2 -\phi^4/12
\end{equation}
We obtain
\begin{equation}
f(\phi) = \mbox{constant} - h \phi - (1-T/T_c)\phi^2/2 + T
\phi^4/(12T_c)
\end{equation}
Note that the model is symmetric with respect to the interchange of
$(h,\phi)$ and $(-h,-\phi)$. This symmetry is approximately valid
for binary liquid mixtures near $T_c$.

For $h=0$ we obtain the order parameter value $\phi_b$ for bulk
coexistence from $df/d\phi=0$.  This leads to $\phi_b = \pm \phi_0$,
with, for $T/T_c \approx 1$,
\begin{equation}
\phi_0^2 = 3(1-T/T_c)
\end{equation}
The free energy near $T_c$ can be rewritten as
\begin{equation}
f(\phi) = \mbox{constant} - h \phi + (\phi^2 - \phi_0^2)^2/12
\end{equation}
In general, $\phi_b$ is found by minimizing $f(\phi)$, and taking
that solution of the cubic equation which has largest modulus. The
other solutions, if real, correspond to metastable and saddle
points. A bulk spinodal results when the latter two coincide.

The Cahn-Landau surface free-energy functional can be written in the
form~\cite{NF,Cahn}
\begin{equation}
\gamma [\phi] =  \int _0 ^{\infty} dz \left \{ \frac{c^2}{4} \left (
\frac {d\phi}{dz} \right ) ^2 + f(\phi(z)) \right \} - h_1 \phi_1 -
g \frac{\phi_1^2}{2}
\end{equation}
where $z \geq 0$ measures the vertical distance from the
liquid-vapour interface downwards into the binary liquid mixture.
The distance is in units of the lattice spacing, or representative
molecular diameter, in the underlying microscopic model. The last
two terms constitute a surface contact energy which depends on the
surface value of the order parameter, $\phi_1 = \phi(z=0)$, and will
be discussed later. We remark that this functional gives only the
surface free energy {\em excess} with respect to a reference value
$\gamma_0$ that is independent of $\phi$, but depends on the
temperature and the material parameters of substrate and adsorbate.
Therefore, the functional $\gamma [\phi]$ does not fulfill, for
example, the positivity requirement of interfacial tensions in
general. Since we will need to calculate only {\em differences} of
surface free energies of substrate/adsorbate configurations, the
unknown $\gamma_0$ drops out.

Having defined the function $f(\phi)$ previously, we now proceed to
relate the coefficient of the gradient squared, $c^2/4$, to
microscopic interaction energies and thermodynamic quantities. This
was done in~\cite{MLI}, with the result
\begin{equation}
c^2/2 = J/k_BT_c \equiv K_c
\end{equation}
For the 3-dimensional Ising model on the SC lattice, $K_c \approx
0.222$, while on the FCC lattice, $K_c \approx 0.102$. For
comparison, the mean-field value for the critical reduced
nearest-neighbour coupling is $K_c = 1/q$, where $q$ is the
coordination number. This gives $K_c = 0.167$ for SC, $K_c = 0.125$
for BCC, and $K_c = 0.083$ for FCC lattices, in the mean-field
theory which we presently employ.

We now address the question whether we can estimate $J$
theoretically. Therefore we turn to the microscopic equation (II.2)
relating $J$ to the molecular pair potentials. The ``mixed" pair
energy $\epsilon_{AM}$ is still to be determined. This is a
non-trivial task, and we limit ourselves here to proposing
reasonable arguments for obtaining a reliable order of magnitude.
This exercise is without quantitative consequences for our theory,
since we will not use this estimate further on, but will take
experimental surface tension data as input for estimating $J$ more
reliably.

To proceed systematically we follow Israelachvili~\cite{Is} and
compare three classes of molecules: non-polar molecules interacting
only through dispersion forces, polar molecules interacting through
dispersion and dipolar forces, and polar molecules interacting also
through hydrogen-bonding in addition to the other forces. For
example, {\em ethane} ($CH_3CH_3$), {\em formaldehyde} ($HCHO$) and
{\em methanol} ($CH_3OH$) have similar size and weight, but are
non-polar, polar, and polar with $H$-bonds, respectively. The
stronger is the interaction, the higher is the liquid-gas critical
temperature of the pure component. For example, for ethane $T_c^{LG}
= 305 K$, for formaldehyde the dipole-dipole interaction leads to
$T_c^{LG} = 408 K$, and additional hydrogen-bonding leads to
$T_c^{LG} = 513 K$ for methanol.

For unlike molecules within the same interaction class, the
Lorentz-Berthelot combining rule~\cite{R},
\begin{equation}
\epsilon_{AB} = \sqrt{\epsilon_{AA} \epsilon _{BB}}
\end{equation}
is often a reasonable approximation. However, for pairs composed of
dissimilar molecules belonging to different classes, this rule can
be a very poor approximation leading to ridiculously low estimates
of the consolute-point temperature, especially when hydrogen-bonding
occurs in one of the two components. The most dramatic manifestation
of this ``non-additivity" of interactions is the hydrophobic effect.
When a non-polar component is mixed with water, the disruption of
the hydrogen-bonded water network is so costly in energy that,
instead of mixing, phase separation is likely to occur. In this case
the mixed interaction strength $\epsilon_{AB}$ is smaller than
either one of the pure values $\epsilon_{AA}$ or $\epsilon_{BB}$,
clearly violating (III.9).

We infer from this that a qualitatively similar phenomenon should
occur when alkanes are mixed with methanol, the latter playing the
role of water in the previous example. The network in this case
consists of {\em one-dimensional chains}~\cite{Is}. The mixed
alkane-methanol interaction lacks the dipole-dipole and
hydrogen-bonding contributions, and to a reasonable approximation
one can say that an alkane molecule sees a methanol molecule as if
it were a molecule of the same weight and size as methanol, but
interacting only through dispersion forces. Therefore, for
determining the mixed pair interaction, we make the rough
approximation to replace methanol effectively by {\em ethane}, and
then we apply the Lorentz-Berthelot rule. In doing so, we also
neglect the dipole/induced-dipole contribution (induction force) to
the mixed-pair van der Waals interaction. However, the induction
force is usually small compared to the dispersion force~\cite{Is}.

For ethane, we estimate $\epsilon_{EE}/k_B = 0.75 T_c^{LG}$, which
gives $229 K$. Using likewise the Lennard-Jones parameters for the
other n-alkanes and adopting the rule (III.9) with ethane playing
the role of methanol, this amounts to the following mixed pair
parameters for alkane-methanol mixtures: $\epsilon_{AM}/k_B = 284 K$
(pentane), $295 K$ (hexane), $ 305 K$ (heptane), $ 313 K$ (octane),
$ 320 K$ (nonane), $ 326 K$ (decane), and $332 K$ (undecane). Now
the nearest-neighbour couplings $J$ can be determined, according to
(II.2), with the results: $J/k_B = 50 K$ (pentane), $52 K$ (hexane),
$53 K$ (heptane), $55 K$ (octane), $56 K$ (nonane), $ 57K$ (decane)
and $ 58 K$ (undecane).

These values can be tested against the reported upper consolute
temperatures $T_c$ (at ambient pressure) for alkane-methanol
mixtures~\cite{Ross1,Ross2}, which are $T_c = 308 K$ (hexane), $324
K$ (heptane), $ 340 K$ (octane), $352 K$ (nonane), $364 K$ (decane)
and $376 K$ (undecane). The ratios $K_c = J/k_B T_c$ are,
respectively, 0.169, 0.164, 0.162, 0.159, 0.157, and 0.154. These
are very reasonable, since they lie in between the $K_c$-values for
SC and FCC packings, and are very close to the mean-field value for
the SC lattice. We conclude that the simple estimates we have made
of the Ising model coupling $J$, based on approximate molecular pair
energies, are consistent with the correct order of magnitude for the
consolute-point temperatures of the binary mixtures and any
reasonable choice of cubic lattice (FCC, BCC or SC) for the lattice
in the model.

A more accurate procedure for determining the parameter $c$, or
equivalently $K_c$, in (III.8) consists of comparing the
experimentally measured liquid-liquid interfacial tension with the
theoretical expression, derived within the Cahn-Landau theory
(e.g.,~\cite{In}),
\begin{equation}
\gamma_{MA} = c \int ^{\phi_0}_{-\phi_0} d\phi \sqrt{f(\phi)}= 2 c
(1-T/T_c)^{3/2}
 \end{equation}
The critical exponent 3/2 is the mean-field value. In real fluids it
is to be replaced by 1.26, and also the amplitude is to be modified.
In order to obtain an estimate for the parameter $c$, we inspect
published interfacial tension data for the nonane/methanol mixture
by Carrillo {\em et al.}~\cite{CTC}, for example. At $T= 298K$ the
measured interfacial tension is $1.47 \times 10^{-3}N/m$.
Alternatively, we can use the data obtained by Kahlweit {\em et
al.}~\cite{KBHJ} for octane and decane and interpolate linearly
between them, which leads to practically the same value $1.45 \times
10^{-3} N/m$. In order to compare this to the dimensionless quantity
$\gamma_{MA}$ of the theory we need to multiply this with the area
of a unit cell in the lattice model, since the distance $z$ is in
these units, and to divide by $k_BT_c$, with $T_c = 352K$.

The lattice constant $\sigma$ is the representative
nearest-neighbour distance of a molecular pair. In the lattice model
$\sigma$ is in principle determined through the number density at
the consolute point, through the relation
\begin{equation}
(\rho_M + \rho_A)\sigma^3 =1,
\end{equation}
where $\rho_{M(A)}$ is the number of methanol (alkane) molecules per
unit volume. While this relation is, strictly speaking, reserved for
$T=T_c$, is actually imposed for all $T$ in the lattice
approximation, since all the cells are filled either by a $M$ or an
$A$ particle. The $\sigma$ defined in this way lies in between the
molecular diameters of the two components, $\sigma_M$ and
$\sigma_A$, as we will illustrate for methanol and nonane in Section
V. In what follows we adopt a simple approximation and just take the
arithmetic mean, $\sigma = (\sigma_M + \sigma_A)/2$.

For methanol $\sigma_M \approx 3.65 \AA$ (valid for both
Lennard-Jones and Stockmayer potentials), while for nonane we may
deduce an effective diameter $\sigma_A \approx 5.99 \AA$ from the
excluded volume or ``hard-sphere" volume. We adopt this useful
approximation even if the molecules are of ellipsoidal rather than
spherical shape~\cite{Is,R}. The relation between the excluded
volume and the associated parameter $b$ in the van der Waals
equation of state is
\begin{equation}
b= 2 \pi \sigma^3/3,\;\;\; \mbox{with}\;\;\; b =
kT^{LG}_c/8P^{LG}_c,
\end{equation}
where $P^{LG}_c$ is the critical pressure of the fluid. Note that if
we use, e.g., the more complicated Peng-Robinson equation of
state~\cite{PR}, a smaller diameter $\sigma_A \approx 5.11 \AA$ is
obtained. This value is almost certainly an underestimation, because
it leads to a molecular mass density of nonane that is higher than
that of methanol, in contradiction with the experimental fact that
the methanol-rich phase is heavier than the nonane-rich phase.

For the van der Waals equation of state we conclude $\sigma \approx
4.83 \AA$, by taking the arithmetic average. The result for the
dimensionless interfacial tension is $\gamma_{MA} = 0.0706$, using
experimental data~\cite{CTC}, and $\gamma_{MA} =0.0696$,
interpolating between experimental data~\cite{KBHJ}. This leads to
$c = 0.587$ and $c=0.579$, and hence $K_c = 0.173$ and $K_c =
0.171$, respectively. This is an interesting result, close to the
values appropriate to the lattices we considered, and very close to
the mean-field value 1/6 for the SC lattice. Alternatively, for the
Peng-Robinson equation of state the results are $K_c = 0.117$ and
$K_c= 0.113$, respectively, which is close to the mean-field value
for the BCC lattice. While being aware of the quantitative
sensitivity of these results to the choice of equation of state, we
expect to obtain a qualitatively correct description by using the
$\sigma$ obtained from the simplest one (van der Waals).

We now turn to the identification of the surface field $h_1$ and the
surface coupling enhancement $g$ in the surface free-energy
functional. For the surface field we have the simple
relation~\cite{MLI},
\begin{equation}
h_1 = H_{surf}/k_BT_c
\end{equation}
where $T_c$ is the consolute-point temperature. For the enhancement
we have, for $T/T_c \approx 1$,
\begin{equation}
g= -m K_c
\end{equation}
with  $m=1$ for the SC lattice~\cite{MLI},and $m=4$ for the FCC
lattice. We conclude that, since $g<0$, first-order as well as
critical wetting transitions are possible, in principle, as was
first demonstrated by Nakanishi and Fisher~\cite{NF}, who derived
the global wetting phase diagrams within Cahn-Landau theory.

\setcounter{equation}{0}
\renewcommand{\theequation}{\thesection.\arabic{equation}}
\section{Crossover from first-order to critical wetting: short-range forces}
The Cahn-Landau theory developed so far does not include the effects
of the  ``tails" of the intermolecular van der Waals forces, but
only takes into account the short-range part of these forces. Such
short-range forces can be constructed artificially, for example,
when the tails of the pair potentials  are ``cut off" at, e.g.,
radial distance $2.5 \sigma$ as is often done in Molecular Dynamics
simulations of fluids~\cite{SIVV}. Our first concern is to check,
within the short-range forces frame-work, the order of the wetting
transition predicted for the n-alkane/methanol mixtures. To this end
we make use of the results derived in~\cite{In}. However, the
notation of that paper cannot be applied directly, because the
definitions of the bulk free energy densities $f$ differ by a factor
12 between that work and the present one (which uses the definitions
of~\cite{MLI}). If we take this into account we find that {\em
tricritical wetting} occurs for
\begin{equation}
\kappa \equiv \sqrt{12} g/c \phi_0 = -2
\end{equation}
Critical wetting takes place for $\kappa < -2$ and first-order
wetting results for $\kappa > -2$ (which includes also $g>0$).

For wetting or drying transitions close to $T_c$ we can use the
previously mentioned expression (III.14) for $g$. If the transition
is not very close to $T_c$ we can use the more general
result~\cite{MLI}, which includes a correction of first order in
$1-T/T_c$, and replace $g$ by $g+(1-T/T_c)/2$. Likewise, for
$\phi_0$ we can use (III.5) or the more generally valid value which
results from solving the equation
\begin{equation}
\phi_0 = \tanh (T_c\phi_0/T)
\end{equation}
Using as input the experimentally determined transition
temperatures, we give the results for $\kappa$, for both methods and
for the two lattices concerned, in Table I.

Experimentally, the hexane/methanol mixture shows a wetting phase
transition, with hexane as the wetting component, at $T_w/T_c
\approx 0.92$. Since in this case the alkane-rich phase wets the
liquid-vapour interface, methanol droplets detach from this
interface. This can be called a ``drying" transition, if we agree
always to refer to methanol as the ``wetting" component. This leads
to $\kappa >- 2$ assuming SC packing and $\kappa < -2$ for FCC, so
that the theory, based on short-range forces alone, would locate
this transition not far from the tricritical point, but probably
still in the critical drying regime. For the heptane/methanol
mixture the drying transition takes place at $T_w/T_c \approx
0.985$, leading to estimates for $\kappa$ well inside the critical
drying range. For octane/methanol no transition was detected.
Possibly it occurs very close to $T_c$. For these three systems no
detailed measurements were made to study the order of the drying
transition.

A clear first-order wetting transition, on the other hand, was
observed for undecane/methanol, with methanol as wetting phase, at
$T_w/T_c \approx 0.903$. The resulting $\kappa$-values satisfy
$\kappa > -2$ (SC) and $\kappa < -2$ (FCC), which indicate that this
transition is not far from the tricritical wetting point $(\kappa =
-2)$ if we neglect the tails of the van der Waals forces. For
decane/methanol $T_w/T_c \approx 0.955$, so that all
$\kappa$-estimates are already within the critical wetting range.
This is more pronounced still for the nonane/methanol mixture, with
$T_w/T_c = 0.992$. For this last mixture short-range critical
wetting is observed experimentally. This suggests that the effect of
the van der Waals forces on this transition is quite weak. The
scrutiny of this is the subject of the next section.

On the basis of the results in Table I we can locate approximately
the tricritical wetting point, corresponding to the choice of the SC
lattice. For both approximations (first and second column of Table
I) we obtain that the tricritical condition $\kappa = -2$ falls
between decane and undecane. Decane/methanol is predicted to show
critical wetting and undecane/methanol to display first-order
wetting. More precisely for the first approximation, valid close to
$T_c$, we obtain $T_w/T_c = 0.917$ for tricriticality. Using linear
interpolation this leads to an effective chain length of 10.72,
which can in principle be achieved by using mixtures~\cite{RBPIM}.
For the second approximation the tricritical point is at $T_w/T_c =
0.941$ and chain length 10.27. Both of these estimates from this
theory with only short-range forces are in qualitative agreement
with our recent experiments which indicate a crossover from
first-order to critical wetting with a tricritical wetting point
between an effective alkane carbon number of 9.6 and
10~\cite{RBPIM}. Since the results for both approximations do not
differ significantly, in contrast to the results for different
choices of lattices, we will henceforth adopt the simple first
approximation valid close to $T_c$, unless stated otherwise.

In the remainder of this section we illustrate typical results based
on the short-range-forces theory, and compute the layer thickness,
the adsorption, and the pertinent surface free energies versus
temperature, at fixed bulk field very close to two-phase
coexistence. We recall that bulk coexistence corresponds to the
condition $h=0$. In view of (II.3-4) and (II.9) this reads $(\mu_M -
\mu_A)+q(\epsilon_{MM}-\epsilon_{AA})/2=0$. Since our system is in a
gravitational field and the chemical potential of a particle (as
defined in the absence of the field) depends on its height through a
gravitational potential energy contribution, this equality is only
satisfied at one particular height, which is, of course, the
position of the {\em liquid-liquid} interface. At a small elevation
$L_e$ above this interface, where the liquid-vapour interface is
situated, there is a non-zero bulk field $h<0$, favouring alkane
molecules. Since we have put all particles on a lattice with a
single lattice constant, the magnitude of $h$ depends not only on
the difference in molecular weight of M and A molecules, but also on
the specific volume per molecule in the real liquid mixture. In
Section V we estimate $h$, give an explicit expression for it, and
show that it is essentially independent of temperature. We can
therefore perform all calculations at fixed $h<0$ for all relevant
$T$, below, at, and above $T_c$.

We will focus on 3 cases: first-order wetting $(\kappa =0$, for
example), tricritical wetting $(\kappa = -2)$ and critical wetting
$(\kappa = - 10$, for example). Before embarking on these cases, we
discuss some generalities concerning the order parameter and the
free energy.

{\bf A. The wetting layer thickness}\\ When a sufficient amount of
methanol is adsorbed at the alkane/vapour interface, it is
convenient to define a layer thickness $l$, which corresponds to the
region occupied by a methanol-rich film. In our model $l$ is
measured from $z=0$ up to the position $z=l$ where the concentration
almost equals that of the alkane-rich phase, which is the bulk
concentration. There is some freedom in defining where precisely
this bulk phase starts, and the results are not sensitive to this
definition, as long as it remains reasonable. We choose to define
$l$ implicitly so that $\phi(l) = 0.9 \phi_b$, which is useful even
for small adsorbed amounts. For thick wetting films on the other
hand, the simpler definition $\phi(l)=0$ would work equally well.
Both types of choices were examined previously when studying alkanes
on water~\cite{IRBBM}. Incidentally, note that the choice
$\phi(l)=\phi_b$ is not possible, since it would result in $ l =
\infty$ in view of the exponential decay of the concentration to its
bulk value.

{\bf B. The adsorption or coverage}\\ The arbitrariness in the
definition of the layer thickness is avoided when working with the
alternative order parameter, the adsorption or coverage. This is a
measure of the total adsorbed amount per unit area, obtained by
integrating the concentration excess. The adsorption $\Gamma$ is
thus defined as
\begin{equation}
\Gamma = \int_0^{\infty} dz (\phi(z)-\phi_b)
\end{equation}
There is a close connection between the adsorption and the
experimentally measured quantity. In the particular case of binary
liquid mixtures that we study, the experiment makes use of
ellipsometry, and the measured ellipticity is proportional to the
adsorption at the liquid-vapour interface~\cite{Bo}, provided the
wetting layer thickness does not exceed $1000\AA$. This condition is
well satisfied in all our experiments. Working with the adsorption
is all the more useful since in the thin-film regime experiments
have been performed with only a small amount of excess material
adsorbed at the interface and in this case a layer thickness is hard
to define.

{\bf C. The surface free energies}\\ If the measurements were
performed precisely at two-phase coexistence of the methanol-rich
and alkane-rich phases, one could work with the three interfacial
tensions $\gamma_{VA}$, $\gamma_{VM}$ and $\gamma_{MA}$, which are
defined as follows. The interfacial tension between methanol and
alkane has already been defined in (III.10), and is an absolute
quantity, since no unknown constant intervenes. The interfacial
tension $\gamma_{VA}$ between the vapour and the alkane-rich phase
can be calculated by minimizing (III.7) with the bulk condition
$\phi \rightarrow \phi_b <0$, and likewise $\gamma_{VM}$ is obtained
using the bulk condition associated with the methanol-rich phase,
$\phi_b > 0$. The latter two surface free energies are relative
quantities, which can be seen most easily from the fact that it
follows from the theory that they are zero at the bulk consolute
point, for a spatially constant order parameter $\phi(z) = 0$. The
true interfacial tension for that hypothetical profile is a
constant, $\gamma_0$, the value of which we do not need to know.

We now define the {\em equilibrium spreading coefficient} $S$,
\begin{equation}
S = \gamma_{VA} - (\gamma_{VM}+ \gamma_{MA})
\end{equation}
Partial wetting corresponds to $S<0$. In this case Young's law
allows us to obtain the contact angle $\theta$ which a methanol-rich
droplet makes against the alkane/vapour interface,
\begin{equation}
\gamma_{VA} = \gamma_{VM} + \gamma_{MA} \cos \theta
\end{equation}
Complete wetting corresponds to an equilibrium spreading coefficient
equal to zero, but if we denote by $\gamma_{VA}^*$ the free energy
of a {\em stable or metastable} surface state with a thin adsorbed
film of the methanol-rich phase, then we can define within
mean-field theory a more general {\em spreading coefficient} $S^*$
through
\begin{equation}
S^* = \gamma_{VA}^* - (\gamma_{VM}+ \gamma_{MA})
\end{equation}
With this definition, $S^*>0$ for the complete wetting regime
between the wetting transition and the upper spinodal point. $S^*$
ceases to be defined for temperatures above this spinodal, as
$\gamma_{VA}^*$ is no longer defined.

A first-order wetting transition is then characterized by a simple
zero-crossing of $S^*$. Critical wetting is more subtle, because in
this case there is no metastable extension of the thin film. $S$
approaches zero from below, with vanishing slope, without crossing
zero. In this case, a generalization $S^*$ different from $S$ does
not exist.

Due to gravitational effects the methanol-rich phase is slightly off
of coexistence, while the alkane-rich phase is stable in bulk, at
the elevation at which the alkane/vapour interface and the methanol
wetting layer reside. Typically, this elevation is 5mm above the
methanol/alkane interface. In practice, while we can calculate under
these circumstances the quantity $\gamma_{VA}$, we must have
recourse to approximations for computing the other two surface free
energies. We will address this interesting problem for the cases of
first-order and critical wetting separately.

\noindent {\bf CASE 1. First-order wetting}\\ When the wetting
transition is of first order, we can, for our present purpose of
illustrating the method, restrict our attention to zero surface
coupling enhancement, $g=0$. Then the surface field $h_1$ which
induces the transition is given by the relation~\cite{In}
\begin{equation}
h_1 = 0.681\, c \phi_{0,w}^2/\sqrt{12},
\end{equation}
where $\phi_{0,w}$ denotes the value of $\phi_0$ and therefore the
temperature at which the wetting transition takes place. We take
$\phi_{0,w}=0.2$ so that $T_w/T_c = 0.987$. The factor $\sqrt{12}$
is present because our units are different from those of~\cite{In}.
Of special interest for us is also the upper spinodal, corresponding
to the metastability limit of the thin film upon increasing the
temperature towards $T_c$. This point is located at
\begin{equation}
\phi_{0,sp}^2 = \sqrt{12}h_1/c,
\end{equation}
where $h_1$ is the surface field defined in (IV.7) and $\phi_{0,sp}$
determines the spinodal temperature. In the experiments the wetting
layers occur at a height where the wetting phase is slightly
undersaturated, corresponding to $h \approx -10^{-6}$ (see Section
V). Here we take a somewhat larger undersaturation, $h = -10^{-5}$,
for greater clarity of presentation. We will encounter the
prewetting transition and the upper prewetting spinodal at slightly
higher temperatures than the wetting transition and associated
spinodal at bulk coexistence. As was mentioned already, the
undersaturation (in methanol) is a consequence of the gravitational
contribution to the chemical potential. The main effect of this is
that the wetting layer cannot reach macroscopic thickness. For a
more detailed study of first-order wetting transitions under
gravity, we refer to reference~\cite{BO}. On the other hand the
lower prewetting spinodal, which is the metastability limit of the
thick film upon decreasing the temperature, typically lies at a
temperature far below that of the equilibrium phase
transition~\cite{B1}.

The phase portrait follows from the first integral or ``constant of
the motion" derived from the Euler-Lagrange equation~\cite{In}
\begin{equation}
\frac{c^2}{2} \frac{d^2\phi}{dz^2} = \frac{df}{d\phi}
\end{equation}
This conservation law reads, with $\dot \phi \equiv d\phi/dz$,
\begin{equation}
\dot \phi = \pm 2 \sqrt{f(\phi)}/c
\end{equation}
While we will work with the SC microscopic lattice model in the
applications to the experiments (Section V), in this methodological
section we work, for a change, with the FCC lattice. We therefore
have $c^2/2 = K_c = 1/12$ in MF theory, implying $c= 1/\sqrt{6}$.
With this choice the value of $h_1$ is fixed by (IV.7) and equals
0.00321. The bulk condition reads
\begin{equation}
\phi(z) \rightarrow \phi_b, \;\;\;\mbox{for}\;\; z \rightarrow
\infty
\end{equation}
which corresponds to the alkane-rich phase. The boundary condition
at the liquid-vapour interface~\cite{In} reads
\begin{equation}
\dot \phi |_{z=0} = - 2 h_1/ c^2
\end{equation}

Figure 1 shows the thickness $l$ of the thin film and the thick film
as a function of temperature, for fixed $h_1$ and $g=0$. The
calculations have been performed for temperatures below as well as
above the consolute temperature $T_c$. For $T > T_c$ there is no
wetting layer, since the bulk phases are fully mixed into a single
phase. However, the preferential adsorption of methanol at the
liquid/vapour interface remains visible as a methanol-enriched
transition zone, which close to $T_c$ is called {\em critical
adsorption}.

We distinguish the thin film and the wetting layer, which, strictly
speaking, is a prewetting layer. The wetting transition (more
precisely, prewetting transition), where the film and the layer
exchange stability, is indicated by the vertical dashed line. The
thin film is stable at low temperature, becomes metastable above the
wetting point and remains metastable up till the spinodal
temperature (open circle). The wetting layer is metastable below the
wetting point and stable above it. When the temperature crosses
$T_c$ the wetting layer state gradually changes to a state in which
a methanol-rich layer sits on a single (fully mixed) bulk phase.

Close to the critical point there is a marked increase in the layer
thickness, below as well as above $T_c$. This is related to the
well-documented phenomenon of critical adsorption~\cite{FdeG}, which
entails a slow (algebraic) decay of the concentration profile into
the bulk phase in place of the usual exponential decay, in this
theory, governed by the length scale set by the bulk correlation
length $\xi$. The divergence of $\xi$ at $T_c$ leads to a diverging
layer thickness, as we have defined it. Since this effect is
sensitive to the definition of the layer thickness, it is preferable
to study the adsorption $\Gamma$, to which we now turn.

Figure 2 shows how the adsorption $\Gamma$ varies with temperature,
for the case of first-order wetting. The vertical line denotes the
wetting transition as in Fig.1. In contrast with the layer
thickness, the adsorption gives a unique and reliable estimate of
the concentration excess near the liquid/vapour interface. In the
thin-film state the adsorption follows closely the layer thickness
variation. However, in the wetting layer the adsorption behaves
qualitatively differently from $l$. Well below $T_c$ the adsorption
depends on the undersaturation, described by the bulk field $h$. If
$h$ is decreased to zero, the wetting layer becomes macroscopically
thick and the adsorption diverges. In mean-field theory this
divergence is logarithmic in $1/|h|$, as our computations confirm.
On the other hand, close to and at $T_c$ the adsorption varies
rapidly as a function of temperature (at fixed undersaturation), as
Fig.2 shows. The value of $\Gamma$ at $T_c$, as a function of $h$,
is described by the scaling laws of critical adsorption~\cite{FdeG}.
In mean-field theory there is (again) a logarithmic divergence in
$1/|h|$, while for real fluids the divergence is of the power-law
form
\begin{equation}
\Gamma  \propto |h|^{(\beta-\nu)/\Delta}
\end{equation}
where $\beta \approx 0.33$ and $\nu \approx 0.63$ are the bulk
order-parameter and correlation-length exponents, respectively, and
$\Delta \approx 1.56$ is the bulk gap exponent which appears when
temperature-like exponents are converted to field-like ones. In
mean-field theory $\beta = \nu = 1/2$ and $\Delta = 1.5$.

The combined result of the increased adsorption at $T_c$ and the
presence of the wetting layer below $T_c$ is what is seen in Fig.2.
We would like to stress that the monotonic behaviour of $\Gamma (T)$
displayed here is not the only possible one. Depending on the
undersaturation, which opposes wetting, and the precise magnitude of
the surface field favouring wetting, non-monotonic adsorption and
critical depletion phenomena can also occur~\cite{MCE}, when the
temperature approaches $T_c$ from above.

Turning now to the surface free energy, Figure 3 shows this quantity
for the thin film and for the wetting layer. The first-order
character of the wetting transition is clearly seen from the
crossing of the free energy branches, and the upper spinodal is also
indicated. Due to the slight undersaturation a spreading coefficient
cannot be defined rigorously, but for small $|h|$ a very good
approximation to $S^*$ defined in (IV.6) is given by
\begin{equation}
S^* \approx \gamma_{thin} - \gamma_{thick}
\end{equation}
Clearly, $S^* <0$ for the thin film, and $S^*>0$ for the wetting
layer. In reality, $S^*$ is slightly larger than this approximation
by an amount of the order of $-h (l_{thick}-l_{thin})$, where $l$ is
the layer thickness. For small $h$, as for our calculations, this
correction is unimportant.

\noindent {\bf CASE 2. Tricritical wetting}\\ The tricritical
wetting transition takes place for the following special values of
the surface coupling enhancement $g$ and surface field
$h_1$~\cite{In},
\begin{equation}
g = - 2\, c \phi_{0,w}/\sqrt{12},
\end{equation}
\begin{equation}
h_1 = -g \phi_{0,w}
\end{equation}
Again we fix the wetting transition at $\phi_{0,w}=0.2$ so that
$T_w/T_c = 0.987$ and we use the same lattice parameters (FCC) as in
the previous case. The computation of the layer thickness $l$ is
straightforward and for a bulk field $h = -10^{-6}$, in the
experimentally relevant range, the result is shown in Figure 4
(curve ``tcw"). The most remarkable feature of this figure is the
weakness of the layer thickness increase in the vicinity of the
wetting transition, at $T/T_c \approx 0.987$. In order to see a
stronger increase of $l$, the bulk field must be made smaller in
magnitude than the $10^{-6}$ we have chosen. However, a smaller
value of $|h|$ will also lead to a larger value of $l$ at $T_c$,
associated with the diverging bulk correlation length at bulk
criticality. With our choice of $h$ we have attempted to visualize
the two effects using the same scale of $l$. The value of $l$ at
$T_c$ is $l_c = 106.3$.

For comparison with experimentally measured quantities, it is more
appropriate to study the adsorption $\Gamma$. The result of the
calculation for the same field $h = -10^{-6}$ is shown in Figure 5
(curve ``tcw"). Now the tricritical wetting transition at $T_w/T_c
\approx 0.987$ is clearly detectable as well as the critical
adsorption phenomenon for $T/T_c$ approaching 1. This is how, in the
theory dealing with short-range forces alone, both phenomena
manifest themselves when the adsorbate is slightly off of two-phase
coexistence in bulk (by fixing $h$). Comparing this with the
adsorption calculated for a first-order wetting transition (and
somewhat larger bulk field magnitude), shown in Fig.2, we can easily
distinguish the continuous and reversible adsorption at tricritical
wetting from the discontinuous and hysteretic behaviour of $\Gamma$
at first-order wetting.

In the experiments it is possible to measure contact angles, and
therefore the spreading coefficient, in spite of the slight
undersaturation of the wetting phase at the liquid-vapour interface.
It would thus be very welcome to be able to calculate $\theta$ or
$S$ for states slightly off of two-phase coexistence, although these
quantities are, strictly speaking, not well defined under these
circumstances. For the case of a first-order wetting transition we
could circumvent this problem by taking advantage of the existence
of two states, the thin film and the wetting layer, so that  $S$
could be approximated as in (IV.14). In the vicinity of continuous
wetting transitions, however, there is only one film state, and we
must have recourse to a new, more general method. In the following
we develop an approximation, which allows us to calculate $S$ for
situations in which the wetting phase is metastable, but
sufficiently long-lived.

For states off of coexistence, only one of the three surface
tensions that feature in the spreading coefficient is well defined.
In our system this is $\gamma_{VA}$, since the alkane-rich phase is
stable in bulk at the height of the liquid-vapour interface, while
the methanol-rich phase is metastable. We denote this metastable
phase by $M^*$. In Figure 6 we have sketched the configuration of a
metastable droplet of methanol attached to the alkane-vapour
interface, for which we would like to calculate $S$ and hence obtain
the contact angle from the Young equation. We begin by formally
defining the following approximation to $S$,
\begin{equation}
S \approx \gamma_{VA} - (\gamma_{VM^*} + \gamma_{M^*A})
\end{equation}
Our task is now to give meaning to and to calculate the last two
terms.

In spite of its thermodynamic metastability the attached droplet is
in mechanical equilibrium, and it takes a long time (typically
weeks) before the droplet disappears, through diffusion, and its
content eventually joins the methanol-rich phase at the bottom.
Clearly, all three interfacial tensions are measurable and should
therefore be calculable, in principle. Their calculation is most
easily explained by examining the phase portrait~\cite{Cahn,IRBBM}
shown in Figure 7. The solid lines with arrows give the
trajectories, which start at points that obey the boundary condition
\begin{equation}
\dot \phi |_{z=0} = - 2 (h_1+ g \phi_1)/ c^2
\end{equation}
and which end at the bulk fixed point, at $\phi_b$, or at the
metastable bulk fixed point at $\phi^*$. The trajectory from
$\phi_1$ to $\phi_b$ is the one that minimizes $\gamma_{VA}$.

We propose to define $\gamma_{VM^*}$ as the excess surface free
energy relative to the metastable bulk phase $M^*$. This excess
quantity is defined using the modified functional
\begin{equation}
\gamma^* [\phi] = \int _0 ^{\infty} dz \left \{ \frac{c^2}{4} \left
( \frac {d\phi}{dz} \right ) ^2 + f^*(\phi(z)) \right \} - h_1
\phi_1 - g \frac{\phi_1^2}{2},
\end{equation}
where $f^*(\phi)$ is the bulk free energy density {\em relative to
the metastable state},
\begin{equation}
f^*(\phi) = -h (\phi-\phi^*) - (1-T/T_c) (\phi^2-\phi^{*2})/2 +
(\phi^4-\phi^{*4})/12
\end{equation}
Minimization of this functional gives the trajectory which runs from
$\phi^*_1$ to $\phi^*$.

Finally, we define the interfacial tension between the metastable
and stable bulk phases $M^*$ and $A$ as follows. Since the
alkane-rich phase is stable in bulk, it is obvious that we must
define $\gamma_{M^*A}$ as the excess free energy relative to the
stable phase, and therefore use $f(\phi)$ defined in (III.6).
Furthermore, the trajectory must start at $\phi^*$ and end at
$\phi_b$. The optimal concentration profile, which minimizes the
surface free energy, does not start with zero derivative, but with a
finite value of $\dot \phi$ (see dotted line in Fig.7). This small
jump in derivative is caused by the slight undersaturation of the
metastable phase, and vanishes for $h \rightarrow 0$. In that limit
$\gamma_{M^*A}$ approaches $\gamma_{MA}$ given by (III.10).

The approximation scheme we adopt neglects the excess free energy in
bulk  of $M^*$ relative to $A$ in calculating $\gamma_{VM^*}$, since
we use $f^*$ in place of $f$. This contribution is to be multiplied
with the thickness, along $z$, of the region occupied by the $M^*$
phase, in order to obtain the excess surface free energy. Hence, our
approach, which is an approximation assuming small undersaturation,
should be more accurate for small droplets than for large ones. Note
that for small undersaturation droplets of a wide range of sizes can
be observed as being metastable.

The result of the approximation is shown in Figure 8. The two
surface free energy curves approach one another almost tangentially,
at $T/T_c \approx 0.987$. At this point the curves actually cross.
However, this cannot be seen on the scale of the figure. The curves
appear coincident for $T/T_c > 0.987$. For example, at $T/T_c =
0.99$ the difference in reduced surface free energy is only about $2
\times 10^{-6}$. The crossing of the curves is due to the fact that
the system is slightly off of coexistence. In the limit $ h
\rightarrow 0$ the curves meet tangentially at the tricritical
wetting point, and the curve associated with $\gamma_{VA}$ stops
there. This is in contrast with Fig.3, where the thin-film state
continues to exist as a metastable state up till the spinodal point.
In fact, tricriticality is achieved when the spinodal coincides with
the wetting transition itself. The difference of the two curves of
Fig.8 gives the spreading coefficient, which will be discussed
together with that for critical wetting.

\noindent {\bf CASE 3: Critical wetting}\\ The critical wetting
transition takes place under the following restrictions of the
surface coupling enhancement $g$ and surface field $h_1$~\cite{In},
\begin{equation}
g < - 2\, c \phi_{0,w}/\sqrt{12},
\end{equation}
\begin{equation}
h_1 = -g \phi_{0,w}
\end{equation}
In this range we choose $g = -10\, c \phi_{0,w}/\sqrt{12}$, and the
wetting temperature, bulk field  and lattice parameter are chosen
the same as for the previous case of tricritical wetting. The result
for the layer thickness $l$ is shown in Fig.4 (curve ``cw"). As in
the previous case, the increase of $l$ in the vicinity of the
critical wetting transition at $T_w/T_c \approx 0.987$ is quite weak
compared with the critical adsorption peak at $T_c$. The value of
$l$ at $T_c$ equals $l_c = 107.3$. Further, the critical wetting
transition presents a weaker signal than the tricritical wetting
transition.

The comparison between tricritical and critical wetting can be made
most clearly in Fig.5, which shows the adsorption. The behaviour at
critical wetting is continuous, while the tricritical transition is
almost discontinuous, which is to be expected in view of the fact
that tricriticality marks the onset of the regime of first-order
(discontinuous) transitions.

The calculation of the surface free energy proceeds along the
approximation scheme outlined in the previous section, and the
result is shown in Figure 9. As in the case of tricritical wetting,
the two curves approach each other almost tangentially at the
wetting point. The difference of the two curves determines the
spreading coefficient $S$, which is shown in Figure 10 for the three
cases: critical wetting (cw), tricritical wetting (tcw), and
first-order wetting (fow). For this comparison, the same bulk field,
$h = -10^{-6}$, was imposed. As expected, $S$ crosses zero linearly
for first-order wetting, while $S$ approaches zero with vanishing
slope for the continuous wetting transitions.

The singular behaviour of $S$ at wetting is described by the power
law, for $T$ approaching $T_w$ from below,
\begin{equation}
S = S_0 (T_w-T)^{2-\alpha_s}
\end{equation}
where $\alpha_s $ is the surface specific heat exponent. At bulk
two-phase coexistence ($h=0$) the Cahn-Landau theory  produces the
mean-field results for short-range forces, $\alpha_s = 1$ (fow),
$\alpha_s = 1/2$ (tcw), and $\alpha_s = 0$ (cw). Since our spreading
coefficients are extensions of $S$ away from two-phase coexistence,
a condition relevant to the experimental situation, it is
instructive to check whether these asymptotic exponents already show
up sufficiently clearly. For ``fow" there is a zero-crossing, so
that $S$ is linear about $T_w$, implying $\alpha_s =1$. For ``tcw"
fits to the calculated curve give $\alpha_s = 0.40 \pm 0.10$ and for
``cw" we find $\alpha_s = 0.02 \pm 0.04$, where the error bars
reflect how much the results typically vary when various temperature
ranges of input data are used. In conclusion, the tricritical and
critical wetting transitions at $h=0$ are already well approximated
by the behaviour of $S$ slightly off of coexistence. This is valid
for systems with short-range forces. In order to compare the theory
with experiments on real fluids, however, it is indispensable to
include the long-range forces in the description, which is the task
to which we now turn.

\setcounter{equation}{0}
\renewcommand{\theequation}{\thesection.\arabic{equation}}
\section{Crossover from first-order to critical wetting: long-range forces}
In this section we include the long-range tails of the van der Waals
interactions between molecules in the Cahn-Landau description. In
doing so we follow~\cite{IRBBM} and limit ourselves to allowing for
a long-range substrate-adsorbate field $h(z)$ which takes into
account the net effect of the substrate-adsorbate adhesive and
adsorbate-adsorbate cohesive contributions which influence the
thickness of the wetting film. The extended Cahn-Landau free-energy
functional reads
\begin{equation}
\gamma [\phi] =  \int _0 ^{\infty} dz \left \{ \frac{c^2}{4} \left (
\frac {d\phi}{dz} \right ) ^2 + f(\phi(z))  \right \} - \int
_{z^*}^{\infty} dz \,h(z)\phi(z) - h_1 \phi_1 - g \frac{\phi_1^2}{2}
\end{equation}
For non-retarded van der Waals forces, relevant to wetting film
thicknesses not exceeding a few hundred $\AA$, the decay of $h(z)$
is algebraic and of the form
\begin{equation}
h(z) = a_3/z^3 + {\cal O}(1/z^4)
\end{equation}
Since in our type of system the long-range interactions favour
wetting by the methanol-rich phase~\cite{deG2}, we have $a_3>0$,
which is referred to as ``agonistic" long-range forces
(LRF)~\cite{deG1}. The leading amplitude $a_3$, to which we will
henceforth refer as LRF amplitude can be related to the Hamaker
constant, which is proportional to the leading term in the
long-range interaction free energy per unit area between the
interfaces that bound the wetting layer. This free energy or {\em
interface potential} $V(l)$, for large wetting layer thickness $l$,
is given by
\begin{equation} V(l) - V(\infty) \approx 2 \phi_0
\int_l^{\infty} dz \,h(z)
\end{equation}
Clearly this potential implies a repulsive force for $a_3>0$. This
force per unit area, or disjoining pressure between the two
interfaces, is
\begin{equation}
\Pi (l) \equiv - dV(l)/dl \approx 2 \phi_0 a_3/l^3
\end{equation}

In the model we will neglect all higher-order contributions to
$h(z)$. This is meaningful when $a_3$ has no significant temperature
dependence, but would not be sufficient for systems in which the
Hamaker constant changes sign, necessitating the inclusion of at
least two terms in $h(z)$ for describing long-range critical
wetting~\cite{IRBBM,RMBIB}.

The long-range field is ``switched on" starting at a cut-off
distance $z^*$. Previous work has devoted special attention to the
possibility of optimizing this parameter~\cite{IRBBM}, but we will
adopt the simplest possible criterion and fix $z^*$ to $2.5 \sigma$,
which is the standard cut-off used in many works.

The basic characteristic of our approach is to take the LRF
amplitude $a_3$ to be an {\em adjustable parameter}, since we ignore
all higher-order terms. If we would consider keeping the full $h(z)$
we could determine $a_3$ by matching it to the Hamaker constant
determined on the basis of experimental data. In spite of the
existence of a theoretical framework for calculating $a_3$ and
higher-order terms~\cite{DL,GD,DN}, we do not embark on this here in
view of the sensitivity of these terms to small changes in molecular
parameters and other microscopic quantities. In particular, the
amplitude $a_4$, i.e. the coefficient of $z^{-4}$ in (V.2), depends
on the details of the spatial variation of the particle density
profiles in the liquid-vapour interface. Therefore, our LRF
amplitude $a_3$ is taken to be an effective constant, whose
magnitude is unknown, but we assume that its sign is consistent with
that of the Hamaker constant in order to capture at least
qualitatively the correct asymptotics for large $z$. Since a
quantitative determination of $h(z)$ is beyond our scope, our
approach is most meaningful in the sense of a {\em perturbative}
one, in which the LRF are treated as a weak contribution. Our
purpose will thus be to test the influence of weak agonistic LRF on
critical wetting and the cross-over to first-order wetting, in
systems which are slightly away from bulk coexistence.

In order to get a feeling for ``weak LRF" and the associated order
of magnitude of $a_3$, it is necessary to check the value of $a_3$
which follows from the Hamaker constant pertaining to the
experiments. Calculation of this quantity is performed using the
dielectric constants and refractive indices for all phases
involved~\cite{Is,BBSRDPBMI}. For the nonane/methanol system, for
example, this leads to the function
\begin{equation}
\tilde\Pi (L) \approx A(T)/L^3 = a(T) k_BT_c/(l^3 \sigma^3) = \Pi(l)
k_B T_c/\sigma^3,
\end{equation}
where $A(T)$ is an energy, and $L$ is the wetting layer thickness in
$\AA$. Since $A$ is proportional to the (bulk coexistence) order
parameter $\phi_0$, it approaches zero for $T \rightarrow T_c$. At
$T_w/T_c = 0.992$, with $T_c = 352 K$, the value $A(T_w)/k_B\approx
3.5 K$ is obtained, so that $a(T_w) \approx 0.010$. In order to
extract $a_3$ we need to calculate also $\phi_0$. This can be done
using (IV.2) at $T_w$ which leads to $\phi_0 = 0.164$. We thus
obtain $a_3 = a/2\phi_0 \approx 0.030$. We remark that $a_3$ can be
considered to be a constant, independent of $T$. In the following
the LRF amplitudes for our calculations will be denoted by ``weak"
provided they are small compared to this estimate.

The remainder of this section is structured as follows. For the
three different mixtures we estimate the bulk and surface fields,
the surface coupling enhancement and calculate the adsorption as a
function of temperature. We interpret the results on the basis of
the knowledge of the properties of the short-range theory
(especially the order and location in temperature of the wetting
transition) and the effect of adding weak long-range forces. We also
compute the specific heat exponent when appropriate. Adsorption
curves and critical exponents are compared with the experimental
results.

{\bf A. Nonane/methanol}\\ The bulk field for this system, which
reflects the difference between the gravitational potential energy
at the liquid-vapour interface and that at the liquid-liquid
interface can be obtained as follows. Equating the
free-energy-density difference between the two bulk phases, due to
the presence of a small bulk field, to the gravitational potential
energy difference per unit volume we obtain the relation
\begin{equation}
2 h \phi_0 k_B T_c/\sigma^3 = -\Delta \rho_{mass} g_m L_e,
\end{equation}
where $g_m$ is the gravitational acceleration and $L_e$ the vertical
distance between the two interfaces, which is roughly $0.5$cm in the
experiments.  The mass density $\rho_{mass}$ in a given phase
involves the molecular weights $m$ of the pure components and the
number densities $\rho$ in the manner
\begin{equation}
\rho_{mass} = m_M \rho_M + m_A \rho_A,
\end{equation}
where $m_M = 32.04 $ amu and for nonane $m_A = 128.25 $ amu (1 amu =
1.66 $\times 10^{-24} g$). The mass density difference $\Delta
\rho_{mass}$, defined as the mass density of the methanol-rich phase
minus that of the alkane-rich phase, is positive.

Our first concern is to obtain a reliable order-of-magnitude
estimate of $h$, based on experimental data. The measured mass
density difference at $T_w$ is $\Delta \rho_{mass} = 0.0216 g/cm^3$.
We obtained this using the same method and apparatus as was used by
Chaar {\em et al.}~\cite{CTC}. Taking this value together with
$\phi_0 = 0.164$, calculated using (IV.2), and invoking the average
diameter $\sigma = 4.83 \AA$ we find $h = -0.873 \times 10^{-6}$.
Note that this  is of the same order of magnitude as the bulk field
assumed in our examples in the short-range theory in the previous
section. An error of $10\%$ in our estimate of $\sigma$ would modify
$h$ by about $30\%$. Also recall that $h<0$, as it should be in
order to stabilize the alkane-rich bulk phase at the height of the
liquid-vapour interface.

In our lattice-gas approximation the mass density difference $\Delta
\rho_{mass}$ can be related to the concentration difference $\Delta
x_M$ of one of the components, using the crude approximation
(III.11) according to which the total number density is constant,
and we obtain
\begin{equation}
\Delta \rho_{mass} \approx (m_M- m_A) \Delta x_M /\sigma^3,
\end{equation}
where $\Delta x_M > 0$ is the concentration of methanol in the
methanol-rich phase minus that in the alkane-rich phase.

This relation relies heavily on the approximation that a single
lattice constant can be employed in the model, and should therefore
not be expected to be accurate. In fact, it ignores the fact that
the heavier alkane molecule occupies in reality a much bigger volume
than the lighter methanol molecule, and therefore gets the sign of
$\Delta \rho_{mass}$ wrong. However, let us not abandon this line of
reasoning yet. Since the bulk order parameter is given by
\begin{equation}
\phi_0 = \Delta x_M
\end{equation}
we can now simplify considerably the expression for the bulk field
and obtain
\begin{equation}
h \approx - g_m L_e (m_M-m_A)/2k_BT_c
\end{equation}
The result is $h \approx 0.806 \times 10^{-6}$. Note that this
approximation, which would be fine if the molecules were of nearly
the same size, accidentally reproduces the correct order of
magnitude of $|h|$. Besides this fortuitous point, the merit of this
simple ``molecular mass"-type of approximation is that it clearly
shows that $h$ is independent of temperature, to the extent that the
height difference $L_e$ between the interfaces is constant.

In order to correct qualitatively for the error made by using a
single effective diameter for the two molecules in the lattice gas
model, we can work with molecular mass densities instead of
molecular masses as follows. Instead of $m_M$ and $m_A$ we employ
effective masses which reproduce to the correct molecular mass
densities when put in the volume $\sigma^3$ of a unit cell in the
lattice. This amounts to the approximation
\begin{equation}
h \approx - g_m L_e (m_M/\sigma_M^3-m_A/\sigma_A^3)\sigma^3/2k_BT_c,
\end{equation}
and leads to the estimate $h = -0.59 \times 10^{-7}$. Now the sign
is correct but the order of magnitude is less satisfactory. Since we
work with alkanes similar to nonane in what follows, and we would
like to exploit the knowledge of the magnitude of $h$ as determined
from experimental input, we will adopt the admittedly heuristic
approximation which consists of using simply (V.10) but with the
correct sign,
\begin{equation}
h \approx - g_m L_e |m_M-m_A|/2k_BT_c
\end{equation}
No qualitative changes are to be expected when using, for example,
(V.11) instead.

The surface field $h_1$ is derived using (II.4) and (II.11). Since
the bulk field contribution in (II.11) is negligible the surface
field is determined by the difference of the pure component
potential parameters and we readily obtain the estimate $ h_1 =
0.0206$.

For estimating the surface coupling enhancement $g$, we follow the
procedure outlined at the end of Section III for determining $K_c$
and then use (III.14). Using the experimental values for the
liquid-liquid interfacial tension and the arithmetic mean $\sigma =
4.83 \AA$ we found $K_c = 0.173$ and $K_c=0.171$. Since these are
close to 1/6 we assume henceforth the simple cubic lattice for
describing this mixture by a lattice model. This leads to the
surface coupling enhancement $g = - K_c = - 1/6 $.

In order to test the sensitivity of $\sigma$ with respect to
alternative ways of defining it, we can use experimental data at the
consolute point~\cite{CTC}, such as mass density $0.689 g/cm^3$,
concentration $x_A=0.29$, partial mass density $m_A \rho_A = 0.62
\rho_{mass}$, and use (III.11) with the result $\sigma = 5.25 \AA$.
This is somewhat larger than the average value we have chosen to
work with, but still well in between the pure component $\sigma$
values.

In the short-range-forces theory the wetting tricritical point for
this system with $g = -1/6$ lies at $\phi_0 = 0.5$ according to
(IV.1), and $h_1= 0.083$ in view of (IV.16). Since the actual
surface field $h_1=0.0206$ is smaller the wetting transition is
critical. This is confirmed by calculating the adsorption in the
short-range forces limit, which leads to a curve very similar to
that for $cw$ in Fig.5. The short-range critical wetting transition
takes place at $\phi_{0,w} = -h_1/g = 0.124$, which corresponds to
$T_w/T_c = 0.995$, quite close to the experimental value of $0.992$.
From (IV.1) we obtain $\kappa = -8.06$. This value is somewhat lower
than the value $-6$ that is given in Table I for n-nonane/methanol
(SC lattice). The reason for this difference is that in the present
section the short-range critical wetting temperature is calculated
self-consistently, whereas in the calculation underlying Table I the
experimental wetting temperatures are used as input.

The influence of weak agonistic long-range forces on this system can
be tested by assuming a LRF amplitude $a_3 = 0.003$, ten times
smaller than the reference value we calculated in the first part of
this section. The prediction from all previous theoretical works is
that the short-range critical wetting transition must become a
first-order wetting transition (see, e.g.,~\cite{deG1,P,NI}).
However, the calculation, represented by the adsorption curve in
Fig.11, clearly reveals a continuous transition, in every respect
reminiscent of the critical wetting phenomenon ($cw$) apparent in
Fig.5. Moreover, the experimentally observed adsorption curve,
through ellipticity measurements, is shown in Fig.12 and is similar
to this theoretical one.

The solution to this paradox lies entirely in the fact that the
system is not at bulk two-phase coexistence. At coexistence the
wetting transition is definitely of first order, by virtue of the
interface potential barrier between the macroscopic (``infinitely"
thick) wetting layer and the thin film. However, off of coexistence
the system does not display first-order wetting but features a
prewetting line, which is very short in temperature as well as in
bulk field, for very weak long-range forces. Under those
circumstances the bulk field due to the gravitational effect is
large enough to make the system sneak {\em underneath} the
prewetting critical point, and show a continuous transition instead
of a (weakly) first-order one. We will clarify this scenario in more
detail for the system decane/methanol in the next subsection. Since
the LRF favour wetting the wetting temperature is slightly lowered
to $T_w/T_c = 0.994$ with respect to the short-range forces limit.
Experimentally, $T_w/T_c= 0.992$ for this system.

We stress that, since we have assumed an arbitrary small $a_3$, we
cannot hope to obtain quantitative agreement with the experiments.
Instead, what we have demonstrated is that slightly away from bulk
coexistence short-range critical wetting can preserve all its
qualitative features when weak long-range forces favouring wetting
are included. If we increase $a_3$ the first-order character of the
wetting transition at bulk coexistence becomes stronger, resulting
in a longer prewetting line off of coexistence. We then find a
first-order prewetting transition instead of a continuous one.

In closing this subsection we calculate the exponent $\alpha_s$
associated with the spreading coefficient for the continuous
prewetting transition, following the new approximation for off-of
coexistence systems developed in Section IV, Cases 2 and 3
(continuous wetting transitions). The result is $\alpha_s = 0.18 \pm
0.1$, which is in fair agreement with the value 0 expected for
short-range critical wetting, and in reasonable agreement with the
experimentally determined value $-0.22 \pm 0.27$.

{\bf B. Decane/methanol}\\ For an alkane chain length of 10 the bulk
and surface field values are changed slightly with respect to the
foregoing case. Using the approximation (V.12) only the molecular
weight of the alkane (for decane $m_A = 142.28 $ amu) and the upper
consolute temperature (for decane/M, $T_c = 364 K$) undergo small
changes, leading to the estimate $h = - 0.893 \times 10^{-6}$ for
decane/methanol. The surface field depends on the difference between
$\epsilon_{AA}/k_B = 464 K$ and $\epsilon^S_{MM}/k_B = 417 K$, where
the superscript S refers to the Stockmayer potential, and we obtain
$h_1 = 0.0323$.

Concerning the surface coupling enhancement, we first estimate the
$K_c$ value from experimental data for the decane/methanol
interfacial tension at ambient temperature. Kahlweit {\em et
al.}~\cite{KBHJ} provided the measured value $1.93 \times 10^{-3}
N/m$, while Carrillo {\em et al.}~\cite{CTC} obtained about $1.71
\times 10^{-3} N/m$. Taking as representative diameter the
arithmetic mean of $\sigma_M = 3.65 \AA$ and $\sigma_A = 6.22 \AA$
based on the van der Waals equation of state with $T_c^{LG}= 617K$
and $P_c^{LG}= 21.1 \times 10^5 Pa$ for decane, we get $\sigma
\approx 4.94 \AA$. This leads to the two estimates for the
dimensionless interfacial tension: $\gamma_{MA} = 0.0938$ and
$0.0831$, respectively. Using (III.10) we obtain $c = 0.607$ and
$0.538$, respectively, and hence $K_c = 0.184$ and $ 0.145$. Since
the average of these two estimates is only 1$\%$ away from 1/6, it
is appropriate to assume also for this system the simple cubic
lattice in the MF model description. We conclude $g = -1/6$ as for
nonane/methanol.

As for the previous mixture, in the short-range-forces theory the
wetting tricritical point for $g = -1/6$ lies at $\phi_0 = 0.5$
according to (IV.1), and $h_1= 0.083$ in view of (IV.16). Since the
surface field $h_1 = 0.0323$ is smaller than this tricritical value,
the predicted short-range wetting transition is critical. It takes
place at $\phi_{0,w} = -h_1/g = 0.194$, which corresponds to
$T_w/T_c = 0.988$. This is larger than the experimentally determined
transition temperature $T_w/T_c = 0.955$, but we have to keep in
mind that agonistic LRF will lower the wetting temperature. From
(IV.1) we obtain $\kappa = -5.15$, which is still well within the
critical wetting regime but closer to triciticality than the
previous mixture. For comparison, we recall that the value of
$\kappa$ predicted by the SRF theory when the experimental wetting
temperature is used as input is also in the critical wetting regime
but  much closer to the tricritical value $-2$. Table I (for decane
and the SC lattice) illustrates this.

The addition of weak long-range forces, for which, for simplicity,
we assume the same strength $a_3 = 0.003$, drives the transition
very weakly first-order in this theory, as is demonstrated by the
remarkable adsorption plot in Fig.13. The wetting temperature is
only slightly lowered to $T_w/T_c = 0.985$, which is not enough to
obtain good agreement with the experimental results. Again, we can
at best hope to get qualitative agreement using the LRF as a weak
perturbation only. The similarity of the adsorption curve of Fig.13
and the typical vertical tricritical wetting adsorption signal
(Fig.5, curve $tcw$) is striking. There is hardly a way to
distinguish the tricritical adsorption jump from a genuine weak
first-order jump of the order parameter. The hysteresis is so minute
that the lower and upper spinodal points $SP_l$ and $SP_u$
practically coincide in temperature.

The physics contained in Fig.13 can be understood in detail by
unraveling the wetting phase diagram associated with these SRF and
LRF parameters for decane/methanol. This is done in Fig.14, which
shows a clear first-order wetting transition at bulk coexistence,
with a first-order prewetting line emerging from it. The lower and
upper spinodal lines merge with this prewetting line at the {\em
prewetting critical point}. For bulk fields larger in magnitude than
the value associated with this point, the transition is
``supercritical" and has the appearance of critical wetting as is
the case in Fig.11 (nonane/methanol). For smaller fields in
magnitude the transition is first-order. In the immediate vicinity
of the prewetting critical point, relevant to decane/methanol in our
approximation of weak LRF forces, the transition appears
tricritical. The adsorption curve of Fig.13 corresponds to the
temperature scan along the dashed line in Fig.14. The approximate
locations of the wetting transition and of the prewetting critical
point are indicated by  open squares.

It is interesting to examine how the prewetting line meets the bulk
coexistence line $h=0$. The two lines meet tangentially~\cite{HS} in
a manner governed by the crossover exponent $\Delta_c$\cite{Ind},
\begin{equation}
h \propto (T-T_w)^{\Delta_c},
\end{equation}
with $\Delta_c = 3/2$ for non-retarded van der Waals forces. Indeed,
our best fit of the numerical data close to $T_w$ leads to the
estimate $\Delta_c = 1.51 \pm 0.01$ and $T_w/T_c \approx 0.98477$.
This clearly shows that the van der Waals tails of the net forces
between interfaces govern the divergence of the wetting layer
thickness and the surface free-energy singularity close to $T_w$ and
for $h \rightarrow 0$. For prewetting in systems with short-range
forces we would have $\Delta_c =1$ and in addition a logarithmic
correction factor would be present.

The experimental adsorption data for decane/methanol are shown in
Fig.15. The data show a rapid continuous rise, but accompanied by
some hysteresis, suggesting that the transition possesses features
of both continuous and first-order character. The obvious presence
of metastability (in the experimental data) provides fairly strong
evidence for an essentially first-order wetting transition.

For the critical exponent $\alpha_s$ our fit to the theoretical
curve of Fig.13 for temperatures below and close to $T_w$ gives
$0.45 \pm 0.1$ which compares favourably with the tricritical value
$1/2$ and less well with the critical value $0$ associated with
prewetting criticality (which is just a MF critical point in our
model). The experimental estimate for this system is $\alpha_s =
0.68 \pm 0.09$. We conclude that although the transition is,
strictly speaking, a first-order transition the exponent $\alpha_s$
is not far from its tricritical value. This is also the case for the
experimental system.
\\
{\bf C. Undecane/methanol}\\ For this system the bulk field
determination proceeds like in the previous ones, taking into
account the molecular weight for undecane $m_A = 156.30 $ amu and
the consolute point for undecane/methanol, $T_c = 376 K$. We obtain
$h = - 0.974 \times 10^{-6}$. For the surface field we use
$\epsilon_{AA}/k_B = 479 K$ for undecane, and get $h_1 = 0.0414$.

Concerning the surface coupling enhancement we examine the published
experimental results for the undecane/methanol interfacial tension,
as obtained by Carrillo {\em et al}~\cite{CTC} at $298 K$. This
gives $2.01 \times 10^{-3} N/m$. Taking as representative diameter
the arithmetic mean of $\sigma_M = 3.65 \AA$ and $\sigma_A = 6.44
\AA$ based on the van der Waals equation of state with $T_c^{LG}=
639K$ and $P_c^{LG}= 19.7 \times 10^5 Pa$ for undecane, we get
$\sigma \approx 5.05 \AA$. The dimensionless interfacial tension
which results after division by $k_BT_c$ and multiplication by
$\sigma^2$ is $\gamma_{MA}= 0.0988$, implying $c= 0.523$ and hence
$K_c = 0.137$. Alternatively, we can interpolate between the data
from Kahlweit {\em et al.}~\cite{KBHJ} for chain lengths 10 and 12
and use the experimental value $2.31 \times 10^{-3} N/m$, or, in
dimensionless units $\gamma_{MA} = 0.1133$, so that $c = 0.600$ and
$K_c = 0.180$. The average of these two values, $K_c \approx 0.158$
is only 5$\%$ away from 1/6, so that also in this case it is
justified to assume the SC lattice to work with, and to set once
again $g= -1/6$ for the surface coupling enhancement.

As for the previous mixtures, the short-range-forces theory places
the wetting tricritical point for $g = -1/6$ at $\phi_0 = 0.5$
according to (IV.1), and $h_1= 0.083$ in view of (IV.16). Since the
surface field $h_1= 0.0414$ is smaller, the short-range wetting
transition is critical. It takes place at $\phi_{0,w} = -h_1/g =
0.248$, which corresponds to $T_w/T_c = 0.979$. This is larger than
the experimentally determined transition temperature $T_w/T_c =
0.903$, which is consistent with the anticipation that agonistic LRF
will lower the wetting temperature. However, since we take the LRF
into account only perturbatively, we cannot expect that $T_w$ will
be lowered sufficiently to obtain good agreement with experiment.
From (IV.1) we obtain $\kappa = -4.03$, which is still well within
the critical wetting regime but closer to triciticality than the two
previous mixtures. Note that this self-consistent determination of
$\kappa$ differs significantly from the $\kappa$ presented in Table
I for n-undecane/methanol (SC lattice), derived using the
experimental wetting temperature as input. Indeed, while the latter
indicates that the short-range wetting transition is already of
first order ($\kappa > -2$), the current self-consistent calculation
requires the intervention of the long-range forces to drive the
wetting transition first-order.

Weak long-range forces, for which for simplicity we assume the same
strength $a_3 = 0.003$ as for both previous mixtures, already drive
the transition clearly first-order, as is shown through the
adsorption displayed in Fig.16. The wetting temperature is only
slightly lowered to $T_w/T_c = 0.976$, due to the fact that our
approach treats the LRF as a small perturbation only. The adsorption
curve of Fig.16 differs somewhat from the typical first-order
adsorption signal (Fig.2) in that the hysteresis is much smaller.
This is due to the vicinity of the system to the prewetting critical
point. Interestingly, the experimental data for this mixture
displayed in Fig.17 show a clear first-order jump but without
hysteresis, unlike for decane/methanol. The fact that the hysteresis
is unobservably small is in line with recent experiments on other
binary liquid mixtures, such as cyclohexane/$CD_3OD$ with
gravity-thinned wetting layers not thicker than about
$100\AA$~\cite{BBMB}. In contrast, in almost density-matched
systems, like cyclohexane/methanol, with gravity-thinned wetting
layers of about $400\AA$, much larger hysteresis is observed. This
has now been understood by calculating the activation energy for
wetting layer nucleation as a function of the film
thickness~\cite{BBMB}. Therefore, the surprising feature is not the
absence of hysteresis for undecane/methanol, but the presence of it
for decane/methanol!

Finally, obviously for this fairly strong first-order wetting
transition the critical exponent analysis immediately leads to
$2-\alpha_s = 1$, as expected, since the mean-field free-energy
branches cross leading to a corner in the equilibrium free energy
(discontinuous derivative).

\setcounter{equation}{0}
\renewcommand{\theequation}{\thesection.\arabic{equation}}
\section{Conclusions}
In this paper we have been concerned with addressing theoretically a
variety of recently observed wetting phenomena displaying the
richness of the wetting phase diagram proposed by Nakanishi and
Fisher~\cite{NF}, featuring first-order, critical and tricritical
wetting. In addition, we have studied how, for the experimental
systems under consideration, the wetting phase transitions are
modified by taking into account approximately the effect of
long-range forces. Many authors, in particular de Gennes, and Ebner
and Saam, predicted that for long-range forces favouring wetting,
critical wetting transitions should become
first-order~\cite{deG1,ES}. This is not what is observed in the
experiments. Mixtures of alkanes and methanol, with the
methanol-rich phase wetting the liquid-vapour interface, have been
observed to pass from continuous wetting to first-order wetting as
the alkane chain length is increased from 9 till 11. The experiments
are consistent with a tricritical wetting transition occurring at
some intermediate effective chain length between 9.6 and
10~\cite{RBPIM}.

The theoretical description which we have explored here supplements
previous works on similar systems in two respects: we have taken
into account that the systems are {\em slightly away from the
conditions of bulk two-phase coexistence}, and, more importantly, we
have allowed for {\em effects of long-range forces between
interfaces}, arising from the inverse power-law van der Waals forces
between the molecules.

It has been necessary to make these extensions of the theory, simply
because several paradoxes concerning the interpretation of the
experiments were up to now left unresolved. For example, the main
apparent contradiction embodied in the observation of short-range
critical wetting~\cite{Ross1,Ross2} is that {\em van der Waals
forces favouring wetting should drive the transition first-order}.
Why was this not seen in the experiments? The answer that we provide
here, and which is complementary to the points of view defended
in~\cite{Ross1,Ross2}, is that weak long-range forces {\em combined
with} a small ``bulk field" which takes the phases just slightly off
of coexistence, do not alter the appearance of short-range critical
wetting.

We have thus obtained evidence that on the one hand critical wetting
can persist, to any reasonable degree of computational accuracy or
practical observability, in slightly undersaturated systems with
weak long-range forces favouring wetting, while on the other hand
first-order wetting is - of course - the generic phenomenon.
Consequently, it is fundamentally interesting to scrutinize the
cross-over between the two. The experiments have shown~\cite{RBPIM}
that this cross-over is consistent with a short-range tricritical
wetting transition scenario. Here we find that this interpretation
is a good approximation, adequate for all practical purposes, even
in systems which instead of displaying strict tricriticality, show a
very short prewetting line emerging from a first-order wetting
transition at bulk coexistence. The role of the tricritical point in
the short-range forces limit is taken over by that of the prewetting
critical point in the presence of agonistic long-range forces, no
matter how weak. The two are difficult to distinguish in practice,
both as regards the order parameter singularity and concerning the
critical exponent associated with the surface free energy
singularity. This exponent, $\alpha_s$, takes the value $1/2$ for
tricritical wetting and the value $0$ for prewetting criticality.

In sum, systems slightly off of coexistence can behave qualitatively
differently from those at coexistence. The most spectacular example
we have found in this regard is the possibility of a
continuous-looking wetting transition under circumstances in which
the wetting transition at coexistence is always of first order. This
is the case when the long-range forces are agonistic, and
consequently short-range first-order wetting remains first-order,
and short-range critical wetting must turn into a first-order
transition, in principle.

The methods we have employed and the theory we have developed are to
a large extent qualitative, at times heuristic, and eventually
amount to a simple mean-field description of the interacting
many-body system. Why should the mean-field theory be a good
approximation in this case? There are several justifications, the
most decisive of which is that the width of the critical region in
which deviations from mean-field critical behaviour should occur is
far too small to be observable experimentally or in Monte Carlo
simulations for equivalent Ising-like systems with short-range
forces. This is a prediction from the latest sophisticated
functional Renormalization Group theory. The second reason is that,
as soon as van der Waals forces are added to the theory, the upper
critical dimension above which mean-field critical exponents are
valid, is lowered from $d_u = 3$ (short-range forces) to $d_u < 3$.
The third reason is the clarity of the investigation. We have
included the effect of a small bulk field, and of a weak
substrate-adsorbate field with algebraic decay, as complications on
top of an otherwise already rich Cahn-Landau theory. It would not be
instructive to add yet a third complication, in the form of forces
arising from interface capillary wave fluctuations, before a full
understanding of the other two refinements has been achieved.
Moreover, due to the undersaturation of the wetting phase
(gravitational thinning of the wetting layer) the parallel and
transverse correlation lengths of the relevant unbinding interface
cannot get large enough for long-wavelength capillary waves to
develop. So the presence of the bulk field renders capillary-wave
considerations superfluous, at least in this system.

While adhering fully to a mean-field theory, we have also indicated
how one can arrive at the surface free-energy functional starting
from a microscopic lattice model of Ising spin-1 type. Further, we
have aimed at providing reasonable estimates for all the
phenomenological parameters in the theory, starting from microscopic
system constants such as molecular interaction potentials (pair
energies and particle diameters) and molecular weights. In this way
the bulk field, surface field and surface enhancements appropriate
to the different mixtures have been related to more fundamental
variables. It has not been possible to complete this scheme fully
self-consistently, and it has been necessary to use as input
experimentally measured interfacial tensions, for example, for
obtaining the appropriate lattice coordination number in the
microscopic model, leading to the use of a simple cubic lattice.

The important parameter for which we have been unable to provide a
reliable system-specific estimate is the amplitude $a_3$ of the
long-range forces. The reason for this is that the theory requires
the knowledge of the entire function $h(z)$ while only the leading
term, related to the Hamaker constant, is known with reasonable
accuracy for a given system. To resolve this draw-back, we have
opted for a perturbative theory, in which the effect of weak
long-range forces is examined on the wetting transitions dictated by
the theory incorporating short-range forces. We have thus fixed
$a_3$ to a value, one order of magnitude smaller than an estimate
based on the Hamaker constant. This is in line with the experimental
fact that the physics predicted by the theory involving short-range
forces is in good accord with the observed continuous wetting
phenomenon in nonane/methanol~\cite{Ross1,Ross2}. The most
remarkable of our findings is that the long-range forces {\em are}
perturbative. Indeed, the inclusion of weak long-range forces does
{\em not} turn the observable critical wetting transition into a
first-order one. This seemingly contradicts previous theoretical
works which indicated that critical wetting must be ruled out for
agonistic LRF. As we emphasized, this is due to the presence of a
small bulk field, turning macroscopic wetting layers into mesoscopic
ones (only hundreds of $\AA$ thick).

As new technical theoretical advances we have achieved the
computation of generalizations of spreading coefficients for systems
slightly displaced from two-phase coexistence in bulk. This is
useful for continuous wetting transitions, in the vicinity of which
only one surface state is thermodynamically stable. Further, we have
found that the adsorption is on many accounts a useful and
well-defined order parameter, which is easy to interpret, in
contrast with the wetting layer thickness. Using the adsorption the
phenomena of (pre-)wetting and critical adsorption can be viewed on
equal footing (see Figs. 2,5,11,13,16) and comparison with the
experimentally measured ellipticity is directly possible.

All in all, we believe that the present work elucidates the systems
in as far as mean-field theory can describe them. The additions of
deviations from bulk coexistence and the perturbative effect of
long-range forces appear crucial to a better modeling and
understanding of the experiments.

\noindent {\bf Acknowledgments.}\\ We thank Michael Fisher, Peter
Monson, Peter Nightingale, Thanassos Panagiatopoulos, J.I. Siepmann,
and Ben Widom for advice concerning the determination of the mixed
pair energy $\epsilon_{AB}$. We further thank Emanuel Bertrand and
Denis Fenistein for discussions and communicating unpublished
results. We are also grateful to Christopher Boulter, Siegfried
Dietrich and the referees for suggesting improvements of the
presentation. This research is supported by the Inter-University
Attraction Poles and the Concerted Action Programme. A.I.P. is
KULeuven Junior Fellow under contract OF/99/045 and F/01/027. LPS de
l'ENS is UMR 8550 of the CNRS, associated with the Universities
Paris 6 and 7. This research has been supported in part by the
European Community TMR Research Network ``Foam Stability and
Wetting", contract nr FMRX-CT98-0171, and by the NRC/NIST
postdoctoral research program.

\newpage
\begin{tabular}{|c|c|c|c|c|}
\hline mixture & SC (app1) & SC (app2) & FCC (app1) & FCC (app2)
\\ \hline n-hexane/methanol & -2.04 & -1.60 & -5.77 & -5.25 \\ n-heptane/methanol & -4.72
& -4.53 & -13.34 & -13.10 \\ n-nonane/methanol & -6.10 & -5.93 &
-17.25 & -17.01
\\ n-decane/methanol & -2.75 & -2.43 & -7.79 & -7.40 \\
 n-undecane/methanol & -1.82 & -1.33 & -5.16 &
-4.58
\\ \hline
\end{tabular}
\vspace{1cm}

\noindent {\bf TABLE I.} Mean-field values for the reduced surface
coupling enhancement $\kappa$, within Cahn-Landau theory for
short-range forces, based on the experimentally determined wetting
temperatures for each mixture. SC (simple cubic) refers to a packing
of molecules with coordination number 6, and FCC (face-centered
cubic)
 corresponds to dense packing with coordination number 12. App1
 corresponds to the approximations (III.5) and (III.14) valid close
 to $T_c$, while app2 corresponds to using the enhancement
 $g+(1-T/T_c)/2$ and (IV.2).
First-order wetting is predicted for $\kappa
> -2$ and critical wetting for $\kappa < -2$, the tricritical value being $\kappa= -2$.
\newpage
{\bf FIGURE CAPTIONS}\\
\\
{\bf Figure 1.} Layer thickness versus temperature for the case of a
first-order (pre-)wetting transition at $T_w/T_c=0.987$ (dashed
line) in the model with short-range forces. The spinodal point (SP)
marks the metastability limit of the thin film. The wetting layer
thickness displays a sharp maximum at the bulk consolute point,
$T=T_c$.\\
\\
{\bf Figure 2.} Adsorption versus temperature for the case of a
first-order (pre-)wetting transition (dashed line) in the model with
short-range forces. The spinodal point (SP) marks the metastability
limit of the thin film. The phenomenon of critical adsorption is
clearly visible when the temperature is lowered from near-critical
values above $T_c$, through $T_c$.
\\
\\
{\bf Figure 3.} The excess surface free energy per unit area
relative to some unknown common constant, for the thin film and the
wetting layer. The crossing point of the curves indicates the
discontinuous (first-order) (pre-)wetting transition for the model
with short-range forces.\\
\\
{\bf Figure 4.}  Layer thickness versus temperature, slightly off of
bulk coexistence, in the vicinity of a tricritical wetting
transition (tcw) and a critical wetting transition (cw) in the model
with short-range forces, occurring at $T_w/T_c= 0.987$. The
(pre-)wetting layer thickness displays a sharp maximum at the bulk
consolute point, $T=T_c$. Note how weak the wetting signal is
compared to the peak at bulk criticality, with this choice of order
parameter $l$.
\\
\\
{\bf Figure 5.} Adsorption versus temperature, slightly away from
bulk two-phase coexistence,  in the vicinity of a tricritical (tcw)
and critical (cw) wetting transition in the model with short-range
forces.  The wetting signals at $T_w/T_c \approx 0.987$ are
comparable in strength to the critical adsorption phenomena at
$T=T_c$. Note how steep is the wetting singularity for the case of
tricritical wetting.
\\
\\
{\bf Figure 6.} The configuration of our system with vapour (V),
alkane-rich (A) and methanol-rich phases (M), showing the metastable
droplet of the methanol-rich phase, M$^*$, attached to the
alkane-vapour interface at height $z=0$. Not shown is the
(microscopic) thin film of M at the A-V interface. Bulk two-phase
coexistence is achieved at the A-M interface, but not at the A-V
interface, where slight gravitational undersaturation of the A phase
in the M component takes place. The bulk concentration at $z=0$ is
indicated by $\phi_b$ and differs slightly from the coexistence
value $-\phi_0$.\\
\\
{\bf Figure 7.} Phase portrait construction allowing, approximately,
to calculate the interfacial tension between the vapour V and
metastable M$^*$ phases, and between the latter and the stable A
phase. The thick straight line represents the boundary condition.
The dashed lines mark the surface values of $\phi$ at the V-A or
V-M$^*$ interfaces. The thick curves give the trajectories of these
two interfaces. The dotted line and the accompanying arrow indicate,
respectively, the jump in the derivative of $\phi$ at the M$^*$-A
interface, and the starting point of the trajectory which, from this
interface, eventually leads to the bulk A phase. For this figure the
value $h=-0.0001$ was used for the bulk field.
\\
\\
{\bf Figure 8.} Dimensionless surface free energy versus temperature
for a system with short-range forces in the vicinity of a
tricritical wetting transition. The lower curve is for the V-A
interface and the upper one is for the combination of the
V-metastable M and metastable M-A interfaces, computed with the new
approximation scheme. Due to the presence of a very small but
nonzero bulk field $h$ the curves approach each other tangentially
and actually cross at the approximate tricritical wetting point at
$T_w/T_c \approx 0.987$. On the scale of the figure the curves
appear to merge.
\\
\\
{\bf Figure 9.} Dimensionless surface free energy versus temperature
for a system with short-range forces in the vicinity of a critical
wetting transition. The lower curve is for the V-A interface and the
upper one is for the combination of the V-metastable M and
metastable M-A interfaces, computed with the new approximation
scheme. Due to the presence of a very small but nonzero bulk field
$h$ the curves approach each other tangentially and actually cross
at the approximate critical wetting point at $T_w/T_c \approx
0.987$. On the scale of the figure the curves appear to merge.
\\
\\
{\bf Figure 10.} Dimensionless spreading coefficients $S$ for
first-order wetting (fow), tricritical wetting (tcw) and critical
wetting (cw), for systems slightly removed from bulk coexistence by
a small bulk field $h = -10^{-6}$ and for short-range forces. For
first-order wetting $S$ clearly displays a zero-crossing at the
transition, implying $2- \alpha_s =1$. For the continuous
transitions $S$ vanishes with vanishing slope, to a very good
approximation, so that $\alpha_s < 1$. For tricritical wetting
$\alpha_s \approx 1/2$ and for critical wetting $\alpha_s \approx
0$. Precise values are given in the text.\\
\\
{\bf Figure 11.} Adsorption of the methanol-rich phase versus
temperature for the model system representing the nonane/methanol
binary liquid mixture at the liquid-vapour interface. The model
parameters bulk field, surface field and surface coupling
enhancement are given in the text. The amplitude of the long-range
forces favouring wetting has been chosen to be very small, treating
these forces as a weak perturbation. Note the continuous wetting
signal at $T_w/T_c \approx 0.994$ and the critical adsorption
approaching $T_c$ from above.
\\
\\
{\bf Figure 12.} Experimentally measured ellipticity versus
temperature, which is proportional to the adsorption, for the
nonane/methanol mixture. The signal is qualitatively the same as in
the theoretical Figure 11. The small dip in the data between wetting
and critical adsorption is within the range of the experimental
noise.\\
\\
{\bf Figure 13.} Adsorption of the methanol-rich phase versus
temperature for the model system representing the decane/methanol
binary liquid mixture at the liquid-vapour interface. The model
parameters bulk field, surface field and surface coupling
enhancement are given in the text. The amplitude of the long-range
forces favouring wetting has been chosen to be very small, treating
these forces as a weak perturbation. Note the very steep and almost
continuous wetting signal at $T_w/T_c \approx 0.986$ and the
critical adsorption near $T_c$. The (pre-)wetting transition is of
first-order, but very weakly so. The lower and upper spinodal points
are also indicated, together with the small jump (dashed line) of
the equilibrium order parameter. The system is very close to the
prewetting critical point.\\
\\
{\bf Figure 14.} Wetting phase diagram for the decane/methanol
system in the variables bulk field and temperature. The first-order
wetting transition at bulk coexistence ($h=0$) is accompanied by the
prewetting line, ending in the prewetting critical point. The lower
and upper spinodal lines which merge at this point are also shown.
The dashed line gives the temperature scan corresponding to Fig.13,
for the fixed bulk field appropriate to the gravity-induced
undersaturation in this system. It should be stressed that the
prewetting line is very short, both in $h$ and in $T/T_c$.
Incidentally, the short-range critical wetting point lies at $T/T_c
\approx 0.988$ (which is outside the range shown).\\
\\
{\bf Figure 15.}  Experimentally measured ellipticity versus
temperature, which is proportional to the adsorption, for the
decane/methanol mixture. Besides a continuous variation, reminiscent
of critical wetting, also hysteresis has been observed, indicating
the first-order character of the transition. \\
\\
{\bf Figure 16.}  Adsorption of the methanol-rich phase versus
temperature for the model system representing the undecane/methanol
binary liquid mixture at the liquid-vapour interface. The model
parameters bulk field, surface field and surface coupling
enhancement are given in the text. The amplitude of the long-range
forces favouring wetting has been chosen to be very small, treating
these forces as a weak perturbation. Note the clear first-order
(pre)wetting transition at $T_w/T_c\approx 0.976$ and the critical
adsorption at $T_c$. The lower and upper spinodal points are also
indicated, together with the  jump (dashed line) of the equilibrium
order parameter. The hysteresis is quite small.\\
\\
{\bf Figure 17.}  Experimentally measured ellipticity versus
temperature, which is proportional to the adsorption, for the
undecane/methanol mixture. A clear first-order transition is seen.
However, no hysteresis is observed.
\end{document}